\font\bigtenrm=cmr10 scaled\magstep5

\line{\hfill First LANL draft, March 30, 2000}

\bigskip
\centerline{\bf{\bigtenrm Does Quantum Nonlocality
Exist?}}
\bigskip
\centerline{\bf{\bigtenrm Bell's Theorem and the}}
\bigskip \centerline{\bf{\bigtenrm Many-Worlds
Interpretation}}

\medskip

\centerline {by}

\medskip
\centerline {Frank J. Tipler\footnote{$^1$}{e-mail
address: tipler@mailhost.tcs.tulane.edu}}
\centerline {Department of Mathematics and
Department of Physics}
\centerline {Tulane University}
\centerline {New Orleans, Louisiana 70118 USA}
\bigskip

\bigskip
\centerline{\bf Abstract}
\bigskip

\noindent
Quantum nonlocality may be an artifact of the
assumption that observers obey the laws of classical 
mechanics, while observed systems obey quantum
mechanics.  I show that, at least in the case of Bell's
Theorem, locality is restored if observed and observer
are both assumed to obey quantum mechanics, as in the
Many-Worlds Interpretation.  Using the MWI, I shall
show that the apparently ``non-local'' expectation value
for the product of the spins of two widely separated
particles --- the ``quantum'' part of Bell's Theorem ---
is really due to a series of three purely local
measurements.  Thus, experiments confirming
``nonlocality" are actually confirming the MWI.


\vfill\eject

Nonlocality is the standard example of a quantum
mechanical property not present in classical
mechanics.  Many papers are published each year (in
1997, four in PRL alone [1]; in 1998, six in PRL alone
[2]; in 1999, eight in PRL alone [3]; and three in {\it
Nature} [4]) on the subject of``nonlocality,'' many (e.g.,
the papers just cited) showing truly awesome
ingenuity.  The phenomenon of nonlocality was first
discussed in the EPR Experiment [5].  We have two spin
1/2 particles, and the two-particle system is in the
rotationally invariant singlet state with zero total
spin angular momentum.  Thus, if we decide to measure
the particle spins in the up-down direction, we would
write the wave function of such a state as

$$ |\Psi> = {{|\uparrow >_1|\downarrow>_2 - 
 |\downarrow >_1|\uparrow>_2}\over \sqrt2}
\eqno(1)$$

\noindent
where the direction of the arrow denotes the direction
of spin, and the subscript denotes the particle.  If we
decide to measure the particle spins in the left-right
direction, the wave function would be written in a
left-right basis as

$$ |\Psi> = {{|\leftarrow >_1|\rightarrow>_2 - 
 |\rightarrow >_1|\leftarrow>_2}\over \sqrt2}
\eqno(2)$$

Nonlocality arises if and only if we assume that the
measurement of the spin of a particle "collapses the
wave function" from the linear superposition
to {\it either} $|\uparrow >_1|\downarrow>_2$ {\it or} 
$ |\downarrow >_1|\uparrow>_2$ in $(1)$.  {\it If}
such a collapse occurs, then measuring the spin of
particle one would fix the spin of particle two.  The
spin of particle two would be fixed instantaneously,
even if the particles had been allowed to separate to
large distances.  If at the location of particle one, we
make a last minute decision to measure the spin of
particle one in the left-right direction rather than the
up-down direction, then instantaneously the spin of
particle two would be fixed in the opposite direction
as particle one --- {\it if} we assume that $(2)$
collapses at the instant we measure the spin of
particle one.  The mystery of quantum nonlocality lies
in trying to understand how particle two changes ---
instantaneously --- in response to what has happened in
the location of particle one.  

\medskip
There is no mystery.  There is no quantum nonlocality. 
Particle two {\it doesn't} know what has happened to
particle one when its spin is measured.  State
transitions are nice and local in quantum mechanics. 
These statements are true because quantum
mechanics tells us that the wave function does {\it
not} collapse when the state of a system is measured. 
In particular, nonlocality disappears when the
Many-Worlds Interpretation (MWI) [6,7,8,16] is adopted. 
The MWI dispels the mysteries of quantum mechanics. 
D.N. Page has previously shown [9] how the EPR reality
criterion is completely fulfilled by the MWI.  I shall
extend his analysis, and show how the ``quantum'' part
of Bell's Theorem [10], namely the expectation valve
for the product of the spins of the two widely
separated electrons, a quantity generally believed to be
essentially non-local, actually arises from a series of
{\it local} measurements.

\medskip
To see how nonlocality disappears,  let us analyze the
measure of the spins of the two particles from the
Many-Worlds perspective.  Let $M_i(...)$ denote the
initial state of the device which measures the spin of
the $i$th particle.  The ellipsis will denote a
measurement not yet having been performed.  We can
for simplicity assume that the apparatus
is 100\% efficient and that the measurement doesn't
change the spin being measured (putting in a more
realistic efficiency and taking into account the fact
that measurement may affect the spin slightly would
complicate the notation but the conclusions would be
unchanged).  That is, if each particle happens to be in
an eigenstate of spin, a measurement of the $i$th
particle changes the measuring device --- but not the
spin of the particle --- as follows:

$${\cal U}_1M_1(...)|\uparrow>_1 =
M_1(\uparrow)|\uparrow>_1, \qquad
{\cal U}_1M_1(...)|\downarrow>_1 = M_1(\downarrow)
|\downarrow>_1\eqno(3)$$

$${\cal U}_2M_2(...)|\uparrow>_2 = 
M_2(\uparrow)|\uparrow>_2, \qquad {\cal
U}_2M_2(...)|\downarrow>_2 = M_2(\downarrow)
|\downarrow>_2\eqno(4)$$

\noindent
where ${\cal U}_i$ are {\it linear} operators which
generates the change of state in the measurement
apparatus, corresponding to the measurement.

\medskip
In particular, if particle 1 is in an eigenstate of spin
up, and particle 2 is in an eigenstate of spin down,
then the effect of the ${\cal U}_i$'s together is

$${\cal U}_2{\cal U}_1M_1(...)M_2(...)|\uparrow>_1
|\downarrow>_2 
 = M_1(\uparrow)M_2(\downarrow)
|\uparrow>_1|\downarrow>_2 \eqno(5)$$

\noindent
even if particles 1 and 2 are light years apart when
their spin orientations are measured.  Similarly,
the result of measuring the {\it i}th particle in the
eigenstate of spin left would be ${\cal
U}_iM_i(...)|\leftarrow>_i =
M_i(\leftarrow)|\leftarrow>_i$, and for an eigenstate
of spin right ${\cal U}_iM_i(...)|\rightarrow>_i =
M_i(\rightarrow)|\rightarrow>_i$, which will generate
equations for spins left and right analogous to eqs. $(3)
- (5)$.

\medskip
Now consider the effect of a measurement on the two
particle system in the Bohm state, that is, with total
spin zero.  This state is $(1)$ or $(2)$ with respect
to an up/down or left/right basis respectively.  The
result is {\it completely} determined by linearity and
the assumed correct measurements on single electrons
in eigenstates.  For example, the effect of
measurements in which both observers happen to
choose to measure with respect to the up/down basis
is  

$${\cal U}_2{\cal U}_1M_2(...)M_1(...)\left[{{
|\uparrow>_1|\downarrow>_2 -
|\downarrow>_1|\uparrow>_2}\over \sqrt2}\right] =$$

$${\cal U}_2M_2(...)
\left[{M_1(\uparrow)|\uparrow>_1
|\downarrow>_2\over\sqrt2} -   {M_1(\downarrow)
|\downarrow>_1| \uparrow>_2\over\sqrt2} \right]$$

$$= {M_2(\downarrow)M_1(\uparrow)|\uparrow>_1
|\downarrow>_2\over\sqrt2} - {M_2(\uparrow)
M_1(\downarrow) |\downarrow>_1|
\uparrow>_2\over\sqrt2}\eqno(6)$$

It may appear from eqn. (6) that it is the first
measurement to be carried out that determines the
split into the two worlds represented by two terms in
$(6)$.  This is false.  In fact, if the measurements are
carried out at spacetime events which are spacelike
separated, then there is no Lorentz invariant way of
determining which measurement was carried out first. 
At spacelike separation, the measuring operators
${\cal U}_1$ and ${\cal U}_2$ commute, and so we can
equally well perform the measurement of the spins of
the electrons in reverse order and obtain the same
splits:

$${\cal U}_1{\cal U}_2M_1(...)M_2(...)\left[{{
|\uparrow>_1|\downarrow>_2 -
|\downarrow>_1|\uparrow>_2}\over \sqrt2}\right] =$$

$${\cal U}_1M_1(...)
\left[{M_2(\downarrow)|\uparrow>_1
|\downarrow>_2\over\sqrt2} -   {M_2(\uparrow)
|\downarrow>_1| \uparrow>_2\over\sqrt2} \right]$$

$$= {M_1(\uparrow)M_2(\downarrow)|\uparrow>_1
|\downarrow>_2\over\sqrt2} - {M_1(\downarrow)
M_2(\uparrow) |\downarrow>_1|
\uparrow>_2\over\sqrt2}\eqno(7)$$

\noindent
the last line of which is the same as that of (6),
(except for the order of states, which is irrelevant).

\medskip
The effect of measurements in which both observers
happen to choose to measure with respect to the
left/right basis is

$${\cal U}_2{\cal U}_1M_2(...)M_1(...)\left[{{
|\leftarrow>_1|\rightarrow>_2 -
|\rightarrow>_1|\leftarrow>_2}\over \sqrt2}\right] =$$

$${\cal U}_2M_2(...)
\left[{M_1(\leftarrow)|\leftarrow>_1
|\rightarrow>_2\over\sqrt2} -   {M_1(\rightarrow)
|\rightarrow>_1| \leftarrow>_2\over\sqrt2} \right]$$

$$= {M_2(\rightarrow)M_1(\leftarrow)|\leftarrow>_1
|\rightarrow>_2\over\sqrt2} - {M_2(\leftarrow)
M_1(\rightarrow) |\rightarrow>_1|
\leftarrow>_2\over\sqrt2}\eqno(8)$$

A comparison of (6)/(7) with (8) shows that {\it
if} two spacelike-separated observers fortuitously
happen to measure the spins of the two particles in
the same direction --- whatever this same direction
happens to be --- both observers will split into two
distinct worlds, and in each world the observers will
measure opposite spin projections for the electrons. 
But at each event of observation, {\it both} of the two
possible outcomes of the measurement will be obtained.
Locality is preserved, because indeed both outcomes
are obtained in total independence of the outcomes of
the other measurement.  The linearity of the
operators ${\cal U}_1$ and ${\cal U}_2$ forces the
perfect anti-correlation of the spins of the particles in
each world.  Since the singlet state is rotationally
invariant, the same result would be obtained whatever
direction the observers happened to choose to measure
the spins.

\medskip
In the EPR experiment, there is actually a third
measurement: the comparison of the two observations
made by the spatially separated observers.  In
fact, the relative directions of the two spin
measurements have no meaning without this third
measurement.  Once again, it is easily seen that
initialization of this third measurement by the two
previous measurements, plus linearity implies that this
third measurement will confirm the split into two
worlds.  In the usual analysis, this third
measurement is not considered a quantum measurement
at all, because the first measurements are considered
as transferring the data from the quantum to the
classical regime.  But in the MWI, there is no classical
regime; the comparison of the data in two macroscopic
devices is just a much a quantum interaction as the
original setting up of the singlet state.  Furthermore,
this ignored third measurement is actually of crucial
importance: it is performed {\it after} information
about the orientation of the second device has been
carried back to the first device (at a speed less than
light!).  The orientation is coded with correlations of
the spins of both electrons, and these correlations (and
the linearity of all operators) will force the third
measurement to respect the original split.  These
correlations have not been lost, for no measurement
reduces the wave function: the minus sign between the
two worlds is present in all eqns. $(1)$ --- $(8)$.

\medskip
To see explicitly how this third measurement works,
represent the state of the comparison apparatus by
$M_c[(...)_1(...)_2]$, where the first entry measures
the record of the  apparatus measuring the first
particle, and the second entry measures the record of
the  apparatus measuring the second particle.  Thus, the
third measurement acting on eigenstates of the
spin-measurement devices transforms the comparison
apparatus as follows:

$${\cal U}_cM_c[(...)_1(...)_2]M_1(\uparrow) =
M_c[(\uparrow)_1(...)_2]M_1(\uparrow)$$

$${\cal U}_cM_c[(...)_1(...)_2]M_1(\downarrow) =
M_c[(\downarrow)_1(...)_2]M_1(\downarrow)$$

$${\cal U}_cM_c[(...)_1(...)_2]M_2(\uparrow) = 
M_c[(...)_1(\uparrow)_2]M_2(\uparrow)$$

$${\cal U}_cM_c[(...)_1(...)_2]M_2(\downarrow) =
M_c[(...)_1(\downarrow)_2]M_2(\downarrow)$$

\noindent
where for simplicity I have assumed the spins will
be measured in the up or down direction.  Then for the
state $(1)$, the totality of the three measurements
together --- the two measurements of the particle
spins followed by the comparison measurement --- is

$${\cal U}_c{\cal U}_2{\cal U}_1 M_c
[(...)_1(...)_2]M_2(...)M_1(...)\left[{{
|\uparrow>_1|\downarrow>_2 -
|\downarrow>_1|\uparrow>_2}\over \sqrt2}\right] =$$

$$= M_c[(\uparrow)_1(\downarrow)_2]
{M_2(\downarrow)M_1(\uparrow)|\uparrow>_1
|\downarrow>_2\over\sqrt2} -
M_c[(\downarrow)_1(\uparrow)_2]{M_2(\uparrow)
M_1(\downarrow) |\downarrow>_1|
\uparrow>_2\over\sqrt2}$$

Heretofore I have assumed that the two observers have
chosen to measure the spins in the same direction.  For
observers who make the decision of which direction to
measure the spin in the instant before the
measurement, most of the time the two directions will
not be the same.  The experiment could be carried out
by throwing away all observations except those in
which the chosen directions happened to agree within a
predetermined tolerance.  But this would waste most
of the data.  The Aspect-Clauser-Freedman Experiment
[11] is designed to use more of the data by testing
Bell's Inequality for the expectation value of the
product of the spins of the two electrons with the spin
of one electron being measured in direction ${\bf\hat
n}_1$, and the spin of the other in direction ${\bf\hat
n}_2$.  If the spins are measured in units of $\hbar/2$,
the standard QM expectation value for the product is

$$<\Psi |({\bf\hat n}_1\cdot{\bf\sigma}_1)
({\bf\hat n}_2 \cdot{\sigma}_2)|\Psi> =  -{\bf\hat
n}_1\cdot {\bf\hat n}_2\eqno(9)$$

\noindent
where $|\Psi>$ is the singlet state (1)/(2).  In
particular, ${\bf\hat n}_1 = {\bf\hat n}_2$ is the
assumed set-up of the previous discussion.  Since the
MWI shows that local measurements in this case
always gives +1 for one electron and -1 for the other,
the product of the two is always -1 in all worlds, and
thus the expectation value for the product is -1, in
complete agreement with (9).

\medskip
To show how (9) comes about by local measurements
splitting the universe into distinct worlds, I follow
[12] and write the singlet state (1)/(2) with
respect to some basis in the ${\bf\hat n}_1$
direction as

$$|\Psi> = (1/\sqrt2)(|{\bf\hat n}_1, \uparrow>_1 |{\bf\hat
n}_1,\downarrow>_2 - |{\bf\hat n}_1, \downarrow>_1 |{\bf\hat
n}_1, \uparrow>_2) \eqno(10)$$

Let another direction ${\bf\hat n}_2$ be the polar
axis, with $\theta$ the polar angle of ${\bf\hat n}_1$
relative to ${\bf\hat n}_2$.  Without loss of generality,
we can choose the other coordinates so that the
azimuthal angle of ${\bf\hat n}_1$ is zero.  Standard
rotation operators for spinor states then give [12]

$$|{\bf\hat n}_1,\uparrow>_2 = (\cos\theta/2) |{\bf\hat
n}_2, \uparrow>_2 \;+\; (\sin\theta/2)|{\bf\hat n}_2,
\downarrow>_2$$

$$|{\bf\hat n}_1,\downarrow>_2 = -\; (\sin\theta/2)
|{\bf\hat n}_2, \uparrow>_2 \;+
\;(\cos\theta/2)|{\bf\hat n}_2, \downarrow>_2$$

\noindent
which yields

$$|\Psi> = (1/\sqrt2)[\; - \;(\sin\theta/2) |{\bf\hat
n}_1,\uparrow>_1|{\bf\hat n}_2,\uparrow>_2
\;+\;(\cos\theta/2) |{\bf\hat
n}_1,\uparrow>_1|{\bf\hat n}_2,\downarrow>_2 $$
$$-\; (\cos\theta/2) |{\bf\hat
n}_1,\downarrow>_1|{\bf\hat n}_2,\uparrow>_2 
\;-\;(\sin\theta/2) |{\bf\hat
n}_1,\downarrow>_1|{\bf\hat n}_2,\downarrow>_2]
\eqno(11)$$

In other words, if the two devices measure the spins
in arbitrary directions, there will be a split into {\it
four} worlds, one for each possible permutation of the
electron spins.  Just as in the case with ${\bf\hat n}_1
= {\bf\hat n}_2$, normalization of the devices on
eigenstates plus linearity forces the devices to split
into all of these four worlds, which are the only
possible worlds, since each observer must measure the
electron to have spin $+1$ or $-1$.

\medskip
The squares of the coefficients in $(11)$ are
proportional to the number of worlds wherein each
respective possibility occurs, these possibilities
being determined by the chosen experimental
arrangement.  This is most easily seen using Deutsch's
MWI derivation [13,16] of the Born Interpretation (BI). 
DeWitt and Graham [7] originally deduced the BI using
the relative frequency theory of probability [14,15],
and this derivation is open to the standard objections to
the frequency theory [14,15].  Deutsch instead derives
the BI using the Principle of Indifference of the
classical/a priori theory of probability [14]. 
According to the Principle of Indifference, the
probability of an event is the number of times the
event occurs in a collection of equipossible cases
divided by the total number of equipossible cases. 
Thus, the probability is 1/6 that a single die throw
will result in a 5, because there are 6 equipossible
sides that could appear, of which the 5 is exactly 1.

\medskip
Deutsch assumes the Principle of Indifference applies
to any experimental arrangement in which the expansion
of the wave function $|\Psi>$ in terms of the
orthonormal basis vectors of the experiment (the
interpretation basis) give equal coefficients for each
term in the expansion.  For example,  both $(1)$ and
$(2)$ are two such expansions, because in both cases
the coefficients of each of the two terms is
$1/\sqrt2$.  The Principle of Indifference thus says
that each of the two possibilities is equally likely in
either the experimental arrangement $(1)$, or in the
interpretation basis $(2)$.  Equivalently, there are an
equal number of worlds corresponding to each term in
either $(1)$ or $(2)$, since in the MWI ``equally
possible'' means ``equal number of worlds'' (equal
relative to a preset experimental arrangement).

\medskip
Deutsch shows [13,16] that if the squares of the
coefficients in the interpretation basis of an
experiment are rational, then a new experimental
arrangement can be found in which the coefficients are
equal in the new interpretation basis.  Applying the
Principle of Indifference to this new set of
coefficients yields the BI for the coefficients in the
original basis.  Continuity in the Hilbert space of wave
functions yields the BI for irrational coefficients
(although it is a presupposition of the MWI that only
coefficients with rational squares are allowed since
irrational squares would imply an irrational number of
worlds).  In particular, the percentage of worlds with
the value of a given basis vector is given by the square
of the coefficient.

\medskip
The expectation value $(9)$ for the product of the
spins is just the sum of each outcome,  multiplied
respectively by probabilities of each of the four
possible outcomes:

$$ {(+1)(+1)P_{\uparrow\uparrow}} \;+\;
 {(+1)(-1)P_{\uparrow\downarrow}}\;+ \;
 {(-1)(+1)P_{\downarrow\uparrow}} \;+ \;
 {(-1)(-1)P_{\downarrow\downarrow}}
\eqno(12)$$

\noindent
where $P_{\uparrow\downarrow}$ is the relative
number of worlds in which the first electron is
measured spin up, and the second electron spin down,
and similarly for the other $P$'s.  Inserting these
relative numbers --- the squares of the coefficients in
$(11)$ ---  into $(12)$ gives the expectation value:

$$ = {1\over 2}\sin^2\theta/2 -
{1\over2}\cos^2\theta/2 - {1\over2}\cos^2\theta/2 + {1\over 2}\sin^2\theta/2
= - \cos\theta = -{\bf\hat
n}_1\cdot {\bf\hat n}_2\eqno(13)$$

\noindent
which is the quantum expectation value
$(9)$.

\medskip
Once again it is essential to keep in mind the third
measurement that compares the results of the two
measurements of the spins, and by bringing the
correlations between the worlds back to the same
location, defines the relative orientation of the
previous two measurements, and in fact determines
whether there is a twofold or a fourfold split.  The
way the measurement of $(9)$ is actually carried out
in the Aspect-Clauser-Freedman Experiment is to let
$\theta$ be random in any single run, and for the
results of each fixed $\theta$ from a series of runs be
placed in separate bins.  This separation requires the
third measurement, and this local comparison
measurement retains the correlations between the
spins.  The effect of throwing away this correlation
information would be equivalent to averaging over all
$\theta$ in the computation of the expectation value: 
the result is $\int_0^\pi <\Psi |({\bf\hat
n}_1\cdot{\bf\sigma}_1) ({\bf\hat n}_2
\cdot{\sigma}_2)|\Psi>\,d\theta = 0$; i.e., the
measured spin orientations of the two electrons are
completely uncorrelated.  This is what we would
expect if each measurement of the electron spins is
completely local, which in fact they are.  There is no
quantum nonlocality.

\medskip
Bell's results [10,15] lead one to think otherwise.  But
Bell made the tacit assumption that each electron's
wave function is reduced by the measurement of its
spin.  Specifically, he assumed that the first
electron's spin was determined by the measurement
direction ${\bf\hat n}_1$ and the value some local
hidden variable parameters $\lambda_1$: the first
electron's spin is given by a function $A({\bf\hat n}_1,
\lambda_1)$.  The second electron's spin is given by an
analogous function $B({\bf\hat n}_2,
\lambda_2)$, and so the hidden variable expectation
value for the product of the spins would not be $(13)$
but instead

$$ \int\rho(\lambda_1,\lambda_2)A({\bf\hat n}_1,
\lambda_1)B({\bf\hat n}_2,
\lambda_2)\, d\lambda_1\,d\lambda_2\eqno(14)$$

\noindent
where $\rho(\lambda_1,\lambda_2)$ is the joint
probability distribution for the hidden variables.  By
comparing a triple set of directions $({\bf\hat n}_1,
{\bf\hat n}_2, {\bf\hat n}_3)$, Bell derived the
inequality $|P({\bf\hat n}_1, {\bf\hat n}_2) - P({\bf\hat
n}_1, {\bf\hat n}_3)| \leq 1 + P({\bf\hat n}_2,{\bf\hat
n}_3)$, which for certain choices of the triple, is
inconsistent with the quantum mechanical $(9)$; i.e., $
{\bf\hat n}_1 = {\bf\hat n}_2 - {\bf\hat n}_3/|{\bf\hat
n}_2 - {\bf\hat n}_3|$ yields $\sqrt2 \leq 1$ if we
assume the MW result $(9)/(13)$, which is the
quantum part of Bell's Theorem.  

\medskip
But $(14)$ assumes that the spin of each particle is
a {\it function} of ${\bf\hat n}_i$ and $\lambda_i$;
that is, it assumes the spin at a location is {\it
single-valued}.  This is explicitly denied by the MWI,
as one can see by letting $\lambda_i$ be the spatial
coordinates of the ith electron.  Bell's analysis tacitly
assumes that the macroscopic world is a single-valued
world like classical mechanics.

\medskip
The automatic elimination of action at a distance
by the MWI is a powerful argument for the validity of
the MWI, for assuming that both single electrons and
many-atom measuring devices are described by
multivalued quantum states.

\medskip
I thank R. Chiao, D. Deutsch, B. DeWitt, and D.N.
Page for helpful discussions, and M. Millis for
inviting me to speak at a NASA conference where
questions of nonlocality were discussed.

\medskip
\centerline{\bf{References}}

\medskip
\noindent\hangindent=20pt\hangafter=1
\item{[1]} M. Horodecki, et al, Phys. Rev. Lett. {\bf
78}, 574 (1997); A. Zeilinger, et al, PRL {\bf
78}, 3031 (1997); E. Hagley, et al, PRL {\bf
79}, 1 (1997); D. Boschi et al, PRL {\bf 79},
2755 (1997); B. Yurke et al, PRL {\bf
79}, 4941 (1997). 

\noindent\hangindent=20pt\hangafter=1
\item{[2]} M. Lewenstein, et al, Phys. Rev. Lett. {\bf
80}, 2261 (1998), quant-ph/9707043; A. Gilchrist, et
al, PRL {\bf 80}, 3169 (1998); J.-W. Pan, et al, PRL
{\bf 80}, 3891 (1998); F. De Martini, PRL
{\bf 81}, 2842 (1998), quant-ph/9710013; W. Tittle et
al, PRL {\bf 81}, 3563 (1998), A. Zeilinger et al, PRL
{\bf 81}, 5039 (1998), quant-ph/9810080. 

\medskip
\noindent\hangindent=20pt\hangafter=1
\item{[3]} D. Bouwmeester et al, PRL {\bf 82}, 1345
(1999), quant-ph/9810035; K. Banaszek et al, PRL {\bf
82}, 2009 (1999), quant-ph/9910117; A. Bramon et al,
PRL {\bf 83}, 1 (1999), hep-ph/9811406; N. Linden et
al, PRL {\bf 83}, 243 (1999), quant-ph/9902022; R.
Polkinghorne et al, PRL {\bf 83}, 2095 (1999),
quant-ph/9906066; S. Aerts et al, PRL {\bf 83}, 2872
(1999), quant-ph/9912064; A. White et al, PRL {\bf
83}, 3103 (1999), quant-ph/9908081; D.A. Meyer, PRL
{\bf 83}, 3751 (1999).

\medskip
\noindent\hangindent=20pt\hangafter=1
\item{[4]}  D. Bouwmeester et al, Nature {\bf
390}, 575 (1997), {\bf 403}, 515 (2000); C.A. Sackett
et al, Nature {\bf 404}, (2000).

\medskip
\noindent\hangindent=20pt\hangafter=1
\item{[5]} A. Einstein, B. Podolsky, and N. Rosen, Phys.
Rev. {\bf 47}, 777 (1935).

\medskip
\noindent\hangindent=20pt\hangafter=1
\item{[6]} H. Everett, Rev. Mod. Phys. {\bf 29}, 454
(1957).

\medskip
\noindent\hangindent=20pt\hangafter=1
\item{[7]}  B.S. DeWitt and N. Graham, ed., {\it The
Many-Worlds Interpretation of Quantum Mechanics}
(Princeton U.P., Princeton, 1973)

\medskip
\noindent\hangindent=20pt\hangafter=1
\item{[8]} S. Goldstein and D.N. Page, PRL {\bf 74}, 3715
(1995).

\medskip
\noindent\hangindent=20pt\hangafter=1
\item{[9]} D.N. Page, Phys. Lett. {\bf 91A}, 57 (1982).

\medskip
\noindent\hangindent=20pt\hangafter=1
\item{[10]} J.S. Bell, Physics {\bf 1}, 195 (1964).

\medskip
\noindent\hangindent=20pt\hangafter=1
\item{[11]} S.J. Freedman and J.F. Clauser, PRL {\bf 28},
938 (1972);  A. Aspect et al, PRL {\bf 47}, 1804 (1982).

\medskip
\noindent\hangindent=20pt\hangafter=1
\item{[12]} D.M. Greenberger et al, Am. J. Phys. {\bf
58}, 1131 (1990). 

\medskip
\noindent\hangindent=20pt\hangafter=1
\item{[13]} D. Deutsch,  Proc. Roy. Soc. London A
(original preprint 1989), quant-ph/9906015. (B.
DeWitt's outline (ref. [16] below), is shorter and may be
easier to understand).

\medskip
\noindent\hangindent=20pt\hangafter=1
\item{[14]} T.L. Fine, {\it Theories of Probability}
(Academic Press, New York, 1973); R. Weatherford,
{\it Philosophical Foundations of Probability Theory}
(Routledge, London, 1982).

\medskip
\noindent\hangindent=20pt\hangafter=1
\item{[15]} M. Jammer, {\it The Philosophy of Quantum
Mechanics} (Wiley, NY, 1974). 

\medskip
\noindent\hangindent=20pt\hangafter=1
\item{[16]} B.S. DeWitt, Int. J. Modern Phys. {\bf A13},
1881--1916 (1998).

\vfill\eject
\bye